\documentclass[aps,prl,twocolumn,superscriptaddress,nofootinbib,longbibliography]{revtex4-2}

\usepackage[T1]{fontenc}
\usepackage[utf8]{inputenc}
\usepackage{lmodern}
\usepackage{microtype}
\usepackage{amsmath,amssymb,mathtools,amsthm}
\usepackage{bm}
\usepackage{braket}
\usepackage{hyperref}
\usepackage[dvipsnames]{xcolor}
\hypersetup{colorlinks=true,linkcolor=MidnightBlue,citecolor=MidnightBlue,urlcolor=MidnightBlue}

\usepackage{svg}

\usepackage{amsmath}

\usepackage{amsmath} 

\theoremstyle{remark}

\newcommand{\prlsection}[1]{\textit{#1.}---}

\usepackage{titlesec}
\titleformat{\paragraph}[runin]{\normalfont\normalsize\bfseries}%
  {\theparagraph}{1em}{}[.]
\titlespacing*{\paragraph}{0pt}{1.5ex}{1em}

\newcommand{\mc}{\mathcal}

\newcommand{\ve}{\varepsilon}
\newcommand{\ketbra}[2]{\mathinner{|{#1}\rangle\langle{#2}|}}

\usepackage{amsmath,amssymb}  
\usepackage{mathrsfs} 

\begin{document}

\title{Leggett–Garg Inequality Violations Bound Quantum Fisher Information
}

\author{Nick Abboud}
\date{\today}
\email{nka2@illinois.edu}
\affiliation{Illinois Center for Advanced Studies of the Universe\\ Department of Physics, 
University of Illinois at Urbana-Champaign, Urbana, IL 61801, USA}

\author{Yuntao Guan}
\date{\today}
\email{yuntaog2@illinois.edu}
\affiliation{Department of Physics, University of Illinois Urbana-Champaign, Urbana IL 61801, USA\\
Anthony J. Leggett Institute for Condensed Matter Theory,
University of Illinois Urbana-Champaign, Urbana IL 61801, USA}

\author{Barry Bradlyn}
\date{\today}
\email{bbradlyn@illinois.edu}
\affiliation{Department of Physics, University of Illinois Urbana-Champaign, Urbana IL 61801, USA\\
Anthony J. Leggett Institute for Condensed Matter Theory,
University of Illinois Urbana-Champaign, Urbana IL 61801, USA}

\author{Jorge Noronha}
\date{\today}
\email{jn0508@illinois.edu}
\affiliation{Illinois Center for Advanced Studies of the Universe\\ Department of Physics, 
University of Illinois at Urbana-Champaign, Urbana, IL 61801, USA}

\begin{abstract}
We prove that a violation of a Leggett--Garg inequality for bounded observables in stationary pure states and thermal states yields a rigorous lower bound on the quantum Fisher information. 
This turns a \emph{qualitative} foundations test of realism in quantum systems into a \emph{quantitative} witness of useful quantum sensitivity and, in the collective setting, into a lower bound on multipartite entanglement depth in many-body systems.
We further demonstrate that Leggett--Garg violations are constrained by the same spectral moments, susceptibilities, and \(f\)-sum-rule bounds that organize many-body response. Our results show that temporal correlations of a single collective observable can serve as an experimentally accessible witness of many-body quantum coherence, without requiring full state reconstruction.
\end{abstract}

\maketitle

\prlsection{Introduction}A central challenge in many-body physics is to determine when a system exhibits genuinely collective and operationally useful quantum behavior \cite{Horodecki:2009zz, Guhne:2008qic, Pezze:2016nxl, Giovanetti2006, Amico:2007ag}. This issue is usually discussed in two very different ways. On the one hand, Leggett-Garg inequalities (LGI) \cite{Leggett:1985zz} (also referred to as temporal \cite{PhysRevLett.71.3235} Bell inequalities \cite{PhysicsPhysiqueFizika.1.195}) probe whether the time history of an observable can be described by a macrorealistic classical picture, and their violation is therefore usually interpreted as a signature of temporal nonclassicality \cite{Emary:2013wfl}. On the other hand, the quantum Fisher information (QFI) \cite{Helstrom:1969fri, Braunstein:1994zz} quantifies how strongly a quantum state responds to an infinitesimal unitary perturbation and has emerged as a central witness of multipartite entanglement and metrological utility \cite{Paris:2008zgg, Toth:2014msl, Pezze:2007jde, Toth:2012lpv, Hyllus:2012ufd, Hauke:2015knb}. These quantities, however, are usually treated as belonging to different conceptual frameworks.
The LGI was originally viewed as a qualitative foundations test \cite{Leggett:2002ifn}, and LGI violations have been observed experimentally in a variety of platforms, including superconducting circuits \cite{Palacios-Laloy:2010rdz, White:2015xdo}, solid-state spin systems \cite{Waldherr:2011kgt, Knee:2011wwi}, and photonic setups \cite{Goggin:2009rae, 2011NatSR...1..101X}; see Ref.\ \cite{Emary:2013wfl} for a review. On the other hand, the QFI is a quantitative resource measure that is often difficult to access directly in many-body systems \cite{Garttner:2017isj, Sone:2020mhz, Beckey:2020htb, Zhang:2022kcj, Yu:2020gye}, requiring frequency-resolved probes \cite{Hauke:2015knb, Scheie:2021jwo}, measurements of multiple observables \cite{Toth:2009ejw,Muller-Rigat:2023vcy, Apellaniz:2017lgo, Strobel:2014npf}, or single-particle-level control \cite{Rath:2021gvz, Vitale:2023few, Yu:2021lgq}.
This raises a natural question: can temporal nonclassicality itself provide a quantitative witness of useful many-body quantum coherence? In other words, can the way a system fluctuates in time reveal, in a rigorous and experimentally useful manner, its metrological sensitivity and, in collective settings, its multipartite entanglement depth?

In this Letter, we answer this question in the affirmative by deriving a model-independent lower bound on the QFI in terms of the violation of an associated Leggett--Garg inequality. The result applies to bounded observables in stationary pure states and thermal states, and at finite temperature is controlled by an explicit universal function. In the collective setting, it can certify finite multipartite entanglement depth through standard QFI criteria \cite{Toth:2012lpv,Hyllus:2012ufd}. While previous work has explored connections between Leggett--Garg violation and quantum metrology \cite{Frowis:2016dew,Moreira:2017rse,Moreira:2019zty}, no general model-independent relation of the kind derived here was previously known. We illustrate the bound for a single qubit and for the transverse-field Ising model, and we use a genuinely \(N\)-partite entangled GHZ state to exhibit exact saturation of the bound in the regime of Heisenberg scaling and macroscopic quantum coherence. By combining our bound with spectral moments, dynamical susceptibilities, and \(f\)-sum-rule constraints, we further embed Leggett--Garg violations in the standard framework of many-body linear-response theory \cite{Forster1995-hn,Fetter2003-jc}.
Through sequential projective measurements in the dichotomic case \cite{Fritz:2010qzm}, and using weak measurements more generally \cite{Aharonov:1988xu,Aharonov:1990zza,Ritchie1991,Jordan:2006lrf,Williams:2007cnr,
Goggin:2009rae,Palacios-Laloy:2010rdz,White:2015xdo,Emary:2017jtt}, our results provide a direct route to certifying useful many-body quantum coherence and, in collective settings, multipartite entanglement depth from dynamics alone.

\prlsection{Stationary Leggett--Garg inequality}We consider bounded observables \(q\) conventionally normalized to take values in the interval \([-1, 1]\) but not necessarily dichotomic. Measuring \(q\) at times \(t_i\) and \(t_j>t_i\) on many identically prepared copies defines a joint distribution \(p_{ij}(q_i,q_j)\) and the two-time correlator
\begin{equation}
C_{ij}=\sum_{q_i,q_j} q_i q_j\, p_{ij}(q_i,q_j).
\end{equation}
For thermal states and stationary pure states, stationarity implies \(C_{ij}=C(t_j-t_i)\). For three equally spaced times \(t_2-t_1=t_3-t_2=\tau\), the Leggett--Garg inequality \cite{Leggett:1985zz} takes the form
\begin{equation}
\label{eq:LGI}
K(\tau)\equiv 2C(\tau)-C(2\tau)\le 1.
\end{equation}
This follows from two assumptions: \(q\) has a definite value at all times, and measuring \(q\) can be made negligibly invasive. The excess \(K(\tau)-1\) therefore quantifies the breakdown of an intuitive macroscopic
understanding of temporal evolution, 
see \cite{Emary:2013wfl}.

In the quantum setting, \(q\) is represented by a Hermitian operator \(Q\) with operator norm \(\|Q\|\le 1\). For a system with Hamiltonian $H$ in a stationary state \(\rho\), time-resolved weak measurements can reconstruct the symmetrized correlation function \cite{Wang2002}
\begin{equation}
\label{eq:C-tau}
C(\tau)=\frac{1}{2}\,\langle \{Q(\tau),Q\}\rangle ,
\end{equation}
where \(Q(\tau)=e^{iH\tau}Qe^{-iH\tau}\) and \(\langle \mathcal O\rangle=\mathrm{Tr}(\rho\mathcal O)\). If \(Q\) is dichotomic, \(Q^2= 1\), then the same correlator can also be obtained from a sequential projective-measurement protocol \cite{Fritz:2010qzm,sm}.

\prlsection{Quantum Fisher information}Considering the unitary family $e^{-i\lambda Q} \rho e^{i\lambda  Q}$ with respect to a parameter $\lambda$, where $ \rho = \sum_n p_n \ketbra{n}{n}$ is an arbitrary state, the spectral representation of the corresponding QFI is $F_Q= \sum_{n, m} f_{n m}$, where \cite{Helstrom:1976}
\begin{equation}\label{eq:f}
    f_{nm} = 2 \frac{(p_n-p_m)^2}{p_n+p_m}\lvert\bra{n} Q\ket{m}\rvert^2.
\end{equation}
For a pure state $ \rho = \ketbra{\psi}{\psi}$, the QFI reduces to the variance
\begin{align} \label{eq:FQ-pure}
    F_Q = 4 \left( \braket{\psi| Q^2|\psi} - \braket{\psi| Q|\psi}^2 \right).
\end{align}
A large QFI implies, through the quantum Cram\'er--Rao bound, that the corresponding problem of determining the value of $\lambda$ admits an unbiased estimator with small variance \cite{Braunstein:1994zz}. The QFI also provides a lower bound on the depth of entanglement in many-body systems \cite{Toth:2012lpv, Hyllus:2012ufd}. Beyond quantum metrology, the QFI has recently emerged as an important quantity in quantum information geometry~\cite{scandi2023quantum,carollo2020geometry,Bengtsson_Zyczkowski_2006} and condensed-matter physics~\cite{mazza2024quantum,wang2025local,balut2025quantum}, where it encodes the Bures geometry of quantum states~\cite{ji2025density,guan2026exploring} and serves as a sensitive probe of collective fluctuations, quantum criticality, and many-body response \cite{Lambert:2023khs,Hauke:2015knb,Ren2024,zanardi2007ground,chowdhury2026information,balut2025quantuma,balut2026fundamental}. We prove below that a violation of the LGI \eqref{eq:LGI} provides a lower bound on the corresponding $F_Q$ associated with the same observable.

\prlsection{LGI-QFI bound}We derive the bound first for a stationary pure state and then for a thermal state.
First, suppose $ \rho = \ketbra{s}{s}$ is a pure stationary state, where $\{\ket{n}\}$ is an eigenbasis of an arbitrary Hamiltonian $ H$. Then, using Eq.~\eqref{eq:C-tau},
\begin{equation}
    K(\tau) - \braket{s| Q^2|s} = \sum_{m \neq s} \lvert\braket{s |  Q | m}\rvert^2 h(\omega_{ s m} \tau),
\end{equation}
where $\omega_{s m} = E_{s} - E_m$ and $h(x) = 2\cos(x) - \cos(2x) - 1$. Similarly, from Eq.~\eqref{eq:FQ-pure},
\begin{align}
    F_Q = 4\sum_{m \neq s} \lvert\braket{s |  Q | m}\rvert^2,
\end{align}
and the model-independent bound
\begin{align} \label{eq:bound-pure}
    F_Q \ge 8\left [K(\tau) - \braket{s| Q^2 |s}\right]
\end{align}
follows immediately from the fact that $h(x) \le \frac{1}{2}$ for $x\in \mathbb{R}$.

\begin{figure}[t]
    \centering
    \includegraphics[width=\columnwidth]{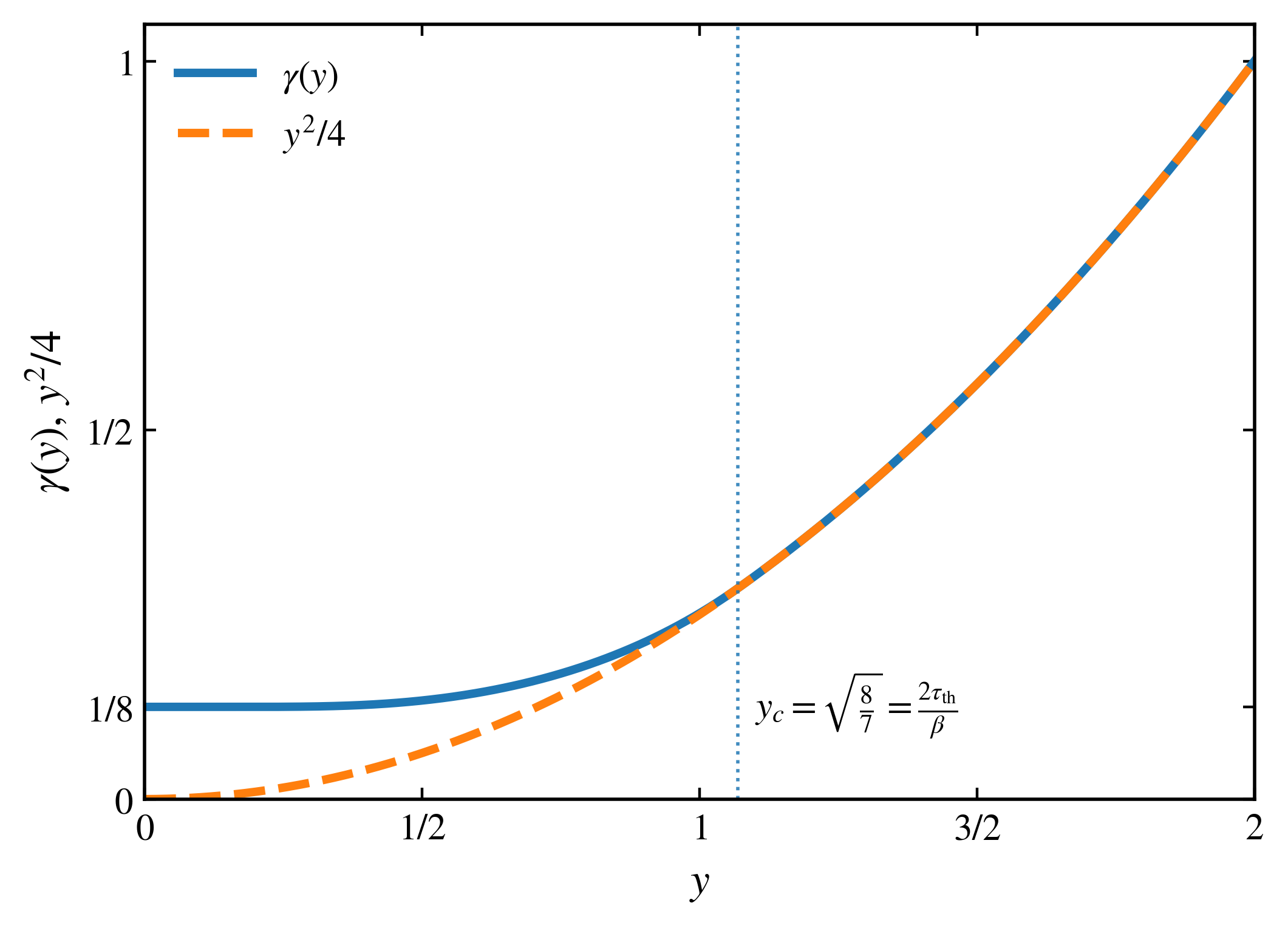}
    \caption{
    Plot of $\gamma(y)$ (solid line) and $y^2/4$ (dashed line) versus
    $y=2\tau/\beta$. Here $\gamma(y)=\max_{x\in\mathbb{R}} R(x,y)$. The vertical dotted line marks the critical value $y_c$. For $2\tau/\beta>y_c$, equivalently $\tau>\tau_{\mathrm{th}}$, one finds $\gamma(y=2\tau/\beta)=\tau^2/\beta^2$.
    }
    \label{fig:gamma_vs_y}
\end{figure}

Next, consider the thermal state $ \rho = \sum_n p_n \ketbra{n}{n}$, where $p_n = e^{-\beta E_n}/Z$, the partition function is $Z = \mathrm{Tr}\,( e^{-\beta H})$, $\beta=1/T$ (with units $k_B=1$), and  $T$ is the temperature. The spectral representation of the Leggett-Garg quantity is now $K(\tau) - \braket{Q^2} = \sum_{n, m} \kappa_{n m}(\tau)$, where
\begin{align} \label{eq:kappa}
    \kappa_{n m}(\tau) = \frac{1}{2}(p_n + p_m) |\braket{n| Q|m}|^2 h(\omega_{nm} \tau).
\end{align}
Comparing Eqs.~\eqref{eq:f} and \eqref{eq:kappa}, we see that $\kappa_{nm}(\tau)= R(\omega_{nm} \tau, 2\tau/\beta)f_{nm}$, where
\begin{align} \label{eq:R}
    R(x, y) &= \frac{1}{4}\coth^2(x/y)h(x).
\end{align}
At fixed $y \neq 0$, the maximum value of Eq.~\eqref{eq:R} is finite, positive, and is achieved for $x \in [0, \pi/2]$. Then, denoting the maximum value by $\gamma(y) = \max_{x\in \mathbb R} R(x, y)$ and observing that $f_{nm} \ge 0$, the term-wise bound $\kappa_{nm} \le \gamma(2\tau/\beta)f_{nm}$ holds, implying
\begin{align} \label{eq:main-result}
    F_Q \ge \frac{K(\tau)-\braket{ Q^2}}{\gamma(2\tau/\beta)}.
\end{align}
We emphasize that the quantity $K(\tau)-\langle Q^2\rangle$ appearing on the right-hand side of the bounds \eqref{eq:main-result} and \eqref{eq:bound-pure} is the difference of correlation functions $2[C(\tau)-C(0)]-[C(2\tau)-C(0)]$, which depends only on the \emph{connected} part of Eq.~\eqref{eq:C-tau}. The weaker bound $F_Q \ge [K(\tau) - 1]/\gamma(2\tau/\beta)$ also holds due to $\lVert  Q \rVert \le 1$.
Importantly, $\gamma$ is a universal function independent of all properties of the system except the temperature $T$ and the measurement time interval $\tau$ (we note also that $\lim_{y\to 0}\gamma(y) = 1/8$, recovering the $T=0$ result). We note also that for $y>\sqrt{8/7}$, $R(x,y)$ achieves its maximum at $x=0$. This leads to the simple form $\gamma(y>\sqrt{8/7})=y^2/4$, see Fig.\ \ref{fig:gamma_vs_y}. Thus, defining the thermal time scale $\tau_{th}=\sqrt{2/7}\beta$, we find that 
\begin{equation}
     F_Q \ge 7\frac{K(z\tau_{th})-\braket{ Q^2}}{2z^2}
\end{equation}
for all $z\geq 1$. We refer the reader to the End Matter for a generalization of Eq.~\eqref{eq:main-result} to an $n$-measurement LGI.

Therefore, our bound shows that if the temporal quantum correlations in a thermal state or stationary pure state are sufficiently nonclassical to violate the LGI, then the state must possess finite distinguishability with respect to perturbations generated by \(Q\), as quantified by the QFI: temporal nonclassicality requires quantitative sensitivity to the same observable. Additionally, our bound implies that the degree of LGI violation is a quantitative  witness of QFI, extending the LGI from a binary qualitative test to a quantitative metric.

\prlsection{Example: thermal qubit}Consider $ H=\frac{1}{2}\ve\sigma_z$, $\varepsilon >0$, so that
\begin{equation}
     \rho = \frac{e^{-\beta \ve/2} \ketbra{\uparrow}{\uparrow} + e^{\beta \ve/2} \ketbra{\downarrow}{\downarrow}}{2\cosh(\beta \ve/2)}.
\end{equation}
The QFI associated with the dichotomic operator $ Q=\hat{\bm n}\cdot {\bm \sigma}$ is $F_Q = 4 \sin^2(\theta) \tanh^2(\beta \ve/2)$, where $\hat{\bm n}\cdot \hat{\bm z} = \cos\theta$ . On the other hand, $K(\tau) = 1 + \sin^2(\theta)h(\ve \tau)$. Combining these expressions, we find
\begin{equation}
    F_Q = \frac{K(\tau) - 1}{R(\ve \tau, 2\tau/\beta)} \ge \frac{K(\tau) - 1}{\gamma(2\tau/\beta)},
\end{equation}
confirming that the bound \eqref{eq:main-result} is satisfied for a qubit in an arbitrary thermal state. If $h(\ve \tau) > 0$, the LGI is violated and the bound is informative.

\prlsection{Example: transverse-field Ising model}In this case, the Hamiltonian for a one-dimensional system with $N$ spins is \cite{sachdev2011}
\(
H=-J\sum_a \sigma_a^z\sigma_{a+1}^z-h\sum_a \sigma_a^x,
\) where $J>0$ is the coupling, $h \neq 0$ is the transverse magnetic field, and
we take \(Q=\sigma_a^z\). At \(T=0\), the QFI is
\(
F_Q=4(1-\langle \sigma_a^z\rangle^2),
\)
so it equals \(4\) for the exact finite-chain ground state and, in the thermodynamic limit, becomes
\(
F_Q=4(1-m^2)
\)
with \(m=(1-h^2/J^2)^{1/8}\) in the ordered phase \(J>h\) \cite{PFEUTY197079}. The corresponding symmetrized correlator \cite{McCoy:1982qx} is even in time and obeys \(|C(\tau)|\le 1\). Its short-time expansion yields
\(
C(\tau)=1-2h^2 \tau^2 +\mathcal O(\tau^4),
\)
and, therefore,
\(
K(\tau)=1+4h^2\tau^2+\mathcal O(\tau^4).
\)
Hence, the LGI is violated at sufficiently early times for \(h \neq 0\), and the result is in agreement with the bound \eqref{eq:bound-pure}.

\prlsection{LGI as multipartite entanglement witness}As a sharp collective benchmark, we consider the GHZ states \cite{Greenberger:1990uox}
\(
\ket{\mathrm{GHZ}_\pm}
=
\frac{1}{\sqrt2}
\left(
\ket{\uparrow\uparrow\cdots\uparrow}
\pm
\ket{\downarrow\downarrow\cdots\downarrow}
\right)
\)
with collective observable
\(
Q =\frac{1}{N}\sum_{a=1}^N \sigma_a^z.
\)
The GHZ states arise as exact eigenstates of \cite{PhysRevLett.107.260502}
\begin{equation}
H_{\rm GHZ}
=
-J\sum_{a=1}^{N-1}\sigma_a^z\sigma_{a+1}^z
+\frac{\Omega}{2}\prod_{a=1}^N \sigma_a^x,
\label{eq:define-GHZ-Hamiltonian}
\end{equation}
$J,\Omega>0$, 
for which the GHZ subspace spanned by
\(
\ket{\uparrow\uparrow\cdots\uparrow}
\)
and
\(
\ket{\downarrow\downarrow\cdots\downarrow}
\)
is invariant. In that subspace, the Hamiltonian reduces to
\(
H_{\text{GHZ,eff}}
=
-J(N-1)I+\frac{\Omega}{2}\tau_z
\)
and $Q=\tau_x$, where $\tau_i$ are Pauli operators in the basis \(\{\ket{\text{GHZ}_+},\ket{\text{GHZ}_-}\}\).
The dynamics therefore become those of an effective qubit and, using $\ket{\text{GHZ}_+}$ as our stationary state, we find
\(
C(\tau) = \cos(\Omega\tau)
\)
and hence
\(
K(\tau)=\left[ 2\cos(\Omega\tau)-\cos(2\Omega\tau)\right],
\)
with maximal LGI violation \(K^{\max}=3/2\) at \(\Omega\tau=\pi/3\). Since \(Q^2=1\) in the GHZ subspace, one directly finds \(
F_Q[Q]=4\) (or, more conventionally \cite{Toth:2014msl}, $F_Q[\widetilde Q]=N^2$, where $\widetilde Q = N Q/2$).
Thus, the GHZ state exactly saturates the LGI--QFI bound \eqref{eq:bound-pure} and provides an explicit example in which Leggett--Garg violation diagnoses Heisenberg scaling, macroscopic coherence, and genuine \(N\)-partite entanglement. See the End Matter for an alternative bound that also certifies $N$-partite entanglement in GHZ states but requires measurements only at times $0$ and $\tau$.

\prlsection{Sum rules, Holevo information, and spectral moments}Our bound can also be combined with standard spectral constraints to obtain nontrivial limitations on Leggett--Garg violations. For thermal states, the QFI admits the response-theory representation \cite{Hauke:2015knb}
\begin{equation}
F_Q
=
-\frac{4}{\pi}\int_0^\infty d\omega \,
\tanh\!\left(\frac{\beta\omega}{2}\right)\chi_{QQ}''(\omega),
\label{eq:FQ_response_repr}
\end{equation}
where $\chi_{QQ}^R(t)
=
-i\,\Theta(t)\,\langle [Q(t),Q(0)]\rangle$ and $\chi_{QQ}''(\omega)
=
\operatorname{Im}\,\chi_{QQ}^R(\omega)$. 
Given that \(\tanh x\le x\) for \(x\ge0\), Ref.\ \cite{Hauke:2015knb} finds the upper bound \(F_Q
\le
-2\beta\int_0^\infty d\omega\,\omega\,\chi_{QQ}''(\omega)/\pi\)
which, when combined with \eqref{eq:main-result}, gives
\begin{align} 
- \frac{2\beta}{\pi}\int_0^\infty d\omega\,\omega\,\chi_{QQ}''(\omega)
\ge   F_Q \ge \frac{K(\tau)-\braket{ Q^2}}{\gamma(2\tau/\beta)}.
\label{eq:LGI_fsum_bound}
\end{align}
Therefore, LGI violations are constrained by the same \(f\)-sum-rule spectral weight that governs linear response. Furthermore, following \cite{Hauke:2015knb}, one can also obtain \(\beta \lim_{\omega\to\infty}\omega^2 \chi'_{QQ}(\omega)
\ge
4\bigl[K(\tau)-\langle Q^2\rangle\bigr]/\gamma(2\tau/\beta)\), 
so the high-frequency tail of the response function also provides an upper bound on LGI violations. Temporal nonclassicality, as measured by the LGI violation, can appear only if the observable \(Q\) coherently connects energy levels, see \eqref{eq:kappa}. The \(f\)-sum rule then limits its magnitude through the finite spectral weight carried by \(Q\). Thus, LGI violations are not only diagnostics of the breakdown of macrorealism \cite{Leggett:1985zz}, but are embedded in the same spectral structure that governs metrological sensitivity and many-body response.

It is also instructive to compare our LGI--QFI bound with the corresponding Holevo-information kernel from Ref.~\cite{mgjv-2p73},
\(
H_{QQ}
=
\int_0^\infty d\omega\, S_{QQ}(\omega)\,\frac{\beta\omega}{e^{\beta \omega}-1}\), with 
\(S_{QQ}(\omega)
=
\sum_{m,n} p_n\, |\langle n|Q|m\rangle|^2\,
\delta\!\left(\omega-\omega_{mn}\right)\).
 The quantity \(H_{QQ}\) quantifies the information about the observable \(Q\) that can leak to an environment purifying the thermal state, characterizing the privacy of correlations in a thermal many-body system. Unlike the QFI kernel, \(\beta\omega/(e^{\beta \omega} - 1)\) is exponentially suppressed at large \(\omega\), so ultraviolet spectral weight can contribute significantly to LGI violation while contributing only weakly to \(H_{QQ}\). For this reason, there is no universal lower bound on \(H_{QQ}\) in terms of LGI violation alone. A lower bound can be recovered only if the spectral support of \(S_{QQ}(\omega)\) is bounded from above, \(\omega\le \Omega^\ast\), in which case the Holevo and LGI kernels are uniformly comparable on the compact interval \([0,\Omega^\ast]\). One may therefore define a finite, support-dependent quantity \(\Gamma_H(\beta,\tau,\Omega^\ast)\) such that
\(
H_{QQ}\ge
\left(K(\tau)-\langle Q^2\rangle\right)/\Gamma_H(\beta,\tau,\Omega^\ast).
\)

This contrast is physically revealing: the QFI admits a universal LGI-based lower bound because its spectral kernel remains sensitive to high-frequency coherent processes, whereas the Holevo quantity is too infrared-weighted for such a universal statement to hold in general. This reflects the fact that LGI violations generally occur at short measurement time $\tau$ and are therefore sensitive to high-frequency spectral information; at long times and low-frequencies, the LGI for thermal states is effectively classical. We note that the bandwidth-limited approximate Holevo bound may play an important role for understanding LGI violation in low-energy effective models. We leave a detailed exploration of this point to future work.

We remark that one can also bound LGI violation using a different type of spectral moment of $\chi''_{QQ}(\omega)$. Let us define at $T=0$ the following quantity 
\(M_2\equiv -(1/\pi) \int_0^\infty d\omega\, \chi''_{QQ}(\omega)\,\omega^2,
\)
which also may be written as $M_2
=
\langle 0| [H,Q]^\dagger [H,Q]|0\rangle
=
\sum_m \lvert\langle 0|Q|m\rangle\rvert^2(E_m-E_0)^2$. 
Since $1-\cos x \leq x^2/2$ for $x\in \mathbb{R}$, one can
obtain the general bound
\(
M_2\ge \left[K(\tau)-\langle Q^2\rangle\right]/\tau^2.
\)
Thus, the violation of LGI in the ground state is also bounded by $M_2$. 

More generally, one may ask whether analogous lower bounds at \(T=0\) can be obtained for higher spectral moments
\(
M_n\equiv - (1/\pi) \int_0^\infty d\omega\,\chi''_{QQ}(\omega)\,\omega^n 
\) for $n>2$.
In contrast with \(M_2\), such bounds are not universal. The obstruction this time is in the infrared: the LGI kernel \(h(\omega\tau)\) vanishes quadratically as \(\omega\to0\), so it controls \(\omega^2\)-weighted spectral weight but not \(\omega^n\)-weighted spectral weight for \(n>2\) near the origin. A lower bound on \(M_n\) with \(n>2\) can therefore be derived only if \(\chi''_{QQ}(\omega)\) has no positive-frequency weight below \(\Delta_{\rm IR}\), in which case
\(
M_n\ge \Delta_{\rm IR}^{\,n-2} M_2
\ge
\Delta_{\rm IR}^{\,n-2}\,
\left(K(\tau)-\langle Q^2\rangle\right)/\tau^2.
\)
Hence, higher-moment bounds require an infrared cutoff, which arises naturally for suitable observables in gapped many-body systems such as insulators where $\Delta_{\rm IR}$ is given by the energy gap at zero temperature.

\prlsection{Conclusion}Our results open a new route to experimentally probing entanglement depth in many-body quantum systems. In contrast to other QFI-based approaches that require frequency resolution, multiple observables, or single-particle control, our approach requires only repeated measurements of a single collective observable. In particular, weak Faraday-rotation measurements in atomic ensembles \cite{Hammerer:2010zjq,Smith:2004twu,Jasperse:2017sdm}, including atoms in optical lattices, can access the required two-time correlations of the normalized collective spin component \(Q=2J_z/N\), where \(J_z=\frac12\sum_{i=1}^N \sigma_i^z\). The corresponding Leggett--Garg combination then yields a rigorous lower bound on the associated QFI and hence on the multipartite entanglement depth in these systems. 
Additionally, recent efforts to measure the QFI spectroscopically in many-body insulators can be reinterpreted through the lens of our bound as placing constraints on nonclassicality in many-body insulators. At the same time, our work shows that the LGI is a witness to quantum geometry.

The universal LGI--QFI bound derived in this work shows that temporal nonclassicality is not merely a qualitative failure of macrorealism, but a quantitative witness of useful many-body quantum coherence. Whenever an observable exhibits sufficiently nonclassical temporal correlations to violate the stationary Leggett--Garg inequality, the same dynamics must also endow the state with finite parameter sensitivity and, in the collective setting, finite multipartite entanglement depth. In this way, our results connect quantum foundations, quantum metrology, and many-body physics through temporal correlations. Therefore, many-body quantum coherence need not be inferred from quantum-state tomography: under suitable conditions, it can also be certified directly from the way a quantum system fluctuates in time.

\begin{acknowledgments}
\prlsection{Acknowledgments}We thank W.~Pfaff for interesting discussions about experimental aspects concerning the bound. We are indebted to the intellectual legacy of Tony Leggett, whose insights across various fields of physics were an inspiration for this work.  The work of Y.G. and B.B. was supported by the U.S. Department of Energy, Office of Basic Energy Sciences, Grant No. DE-SC0026342.   J.N. and N.A.
are partly supported by the U.S. Department of Energy, Office of Science, Office of Nuclear Physics under Award No.
DE-SC0023861. 
\end{acknowledgments}

\bibliography{refs}

\onecolumngrid

\appendix

\section{End Matter}

\prlsection{Bound from the $n$-measurement LGI}Equation \eqref{eq:LGI} is the first in a family of inequalities \cite{Emary:2013wfl}
\begin{align}
    K_p(\tau) \le p-2,\qquad\qquad K_p(\tau) = (p-1)C(\tau) - C([p-1]\tau),
\end{align}
where $p\in \mathbb Z$ and $p \ge 3$, which are satisfied under the same conditions as Eq.~\eqref{eq:LGI}. A straightforward generalization of the arguments leading to Eq.~\eqref{eq:main-result} yields a generalized bound
\begin{align} \label{eq:bound-Kp}
    F_Q \ge \frac{K_p(\tau) - (p-2)\braket{Q^2}}{\gamma_p(2\tau/\beta)},
\end{align}
where
\begin{align}
    \gamma_p(y) = \max_{x \in \mathbb R}\left[ \frac{1}{4} \coth^2(x/y)h_p(x) \right], \qquad \qquad  h_p(x) = (p-1)\cos(x) - \cos([p-1]x) - (p -2).
\end{align}
We note that
\begin{align}
    K_p(\tau) - (p-2)\braket{Q^2} = \frac{1}{2}\sum_{n,m} (p_n + p_m) \lvert\braket{n|Q|m}\rvert^2 h_p(\omega_{nm} \tau) \le \braket{Q^2} h_p^\text{max},
\end{align}
where $h_p^\text{max} = 2 + \mc O(1/p)$. Similarly, for fixed $y$, we find $\gamma_p(y) \sim (p^2y^2/8)[1 + \mc O(1/p)]$. Therefore,
\begin{align}
    \frac{K_p(\tau) - (p-2)\braket{Q^2}}{\gamma_p(2\tau/\beta)} \le \frac{4\beta^2\braket{Q^2}}{p^2 \tau^2}\left[1 + \mc O(1/p)\right].
\end{align}
Therefore, at fixed $\tau$ and $\beta$, the bound \eqref{eq:bound-Kp} becomes uninformative for large $p$. For $p$ of order one, Eq.~\eqref{eq:bound-Kp} can be modestly stronger than Eq.~\eqref{eq:main-result}, but not parametrically so. If instead $(p-1) \tau$ is held fixed as $p \rightarrow \infty$, then $\gamma_p(2\tau/\beta)$ approaches a constant and $K_p(\tau) - (p-2)\braket{Q^2}$ approaches zero, so the bound \eqref{eq:bound-Kp} is also uninformative in this limit.

\prlsection{Certifying $N$-partite entanglement with only two measurement times}The same arguments that led to Eq.~\eqref{eq:main-result} also establish the alternative lower bound
\begin{align} \label{eq:bound-two-time}
    F_Q \ge \frac{\braket{Q^2} - C(\tau)}{\tilde \gamma(2\tau/\beta)} \ge 0, \quad \qquad \tilde \gamma(y) = \max_{x \in \mathbb R}\left[\frac{1}{4} \coth^2(x/y) (1-\cos x)\right].
\end{align}
In the qubit and transverse-field Ising model examples discussed in the main text, Eq.~\eqref{eq:bound-two-time} yields a strictly weaker bound than Eq.~\eqref{eq:bound-pure}. However, in the GHZ example, where $C(\tau) = \cos(\Omega \tau)$, the maximum value of $\braket{Q^2}-C(\tau)$ is $2$ and at zero temperature we have $\lim_{T\rightarrow 0}\tilde \gamma(2\tau/\beta) = 1/2$, so Eq.~\eqref{eq:bound-two-time} yields $F_Q \ge 4$. Thus, Eq.~\eqref{eq:bound-two-time} produces the same bound on $F_Q$ as Eq.~\eqref{eq:bound-pure} in this case and is therefore capable of detecting $N$-partite entanglement in GHZ states. From an experimental perspective, we note that, unlike Eq.~\eqref{eq:main-result}, the bound \eqref{eq:bound-two-time} only requires two measurement times: it requires reconstructing $C(0)=\braket{Q^2}$ and $C(\tau)$, but not $C(2\tau)$.

\end{document}


\title{Leggett–Garg Inequality Violations Bound Quantum Fisher Information\\ Supplemental Material
}

\author{Nick Abboud}
\date{\today}
\email{nka2@illinois.edu}
\affiliation{Illinois Center for Advanced Studies of the Universe\\ Department of Physics, 
University of Illinois at Urbana-Champaign, Urbana, IL 61801, USA}

\author{Yuntao Guan}
\date{\today}
\email{yuntaog2@illinois.edu}
\affiliation{Department of Physics, University of Illinois Urbana-Champaign, Urbana IL 61801, USA\\
Anthony J. Leggett Institute for Condensed Matter Theory,
University of Illinois Urbana-Champaign, Urbana IL 61801, USA}

\author{Barry Bradlyn}
\date{\today}
\email{bbradlyn@illinois.edu}
\affiliation{Department of Physics, University of Illinois Urbana-Champaign, Urbana IL 61801, USA\\
Anthony J. Leggett Institute for Condensed Matter Theory,
University of Illinois Urbana-Champaign, Urbana IL 61801, USA}

\author{Jorge Noronha}
\date{\today}
\email{jn0508@illinois.edu}
\affiliation{Illinois Center for Advanced Studies of the Universe\\ Department of Physics, 
University of Illinois at Urbana-Champaign, Urbana, IL 61801, USA}

\maketitle

In this Supplemental Material, we derive the Leggett-Garg inequality (LGI) used in the main text, and we show how the two-time correlators obtained from the appropriate measurement protocols reproduce the symmetrized correlator $\frac{1}{2}\braket{\{Q(\tau),Q\}}$ in quantum systems, as asserted in the main text. Specifically, in Sec.~\ref{ssec:LGI}, we review a standard derivation \cite{Leggett:1985zz, Emary:2013wfl} of the LGI in Eq.\ (2) of the main text for possibly non-dichotomic but bounded observables. In Sec.~\ref{ssec:C-in-QM-A}, we show that the correlation function of successive projective measurements of $Q$ equals the symmetrized correlator when $Q$ is dichotomic \cite{Fritz:2010qzm}, and in Sec.~\ref{ssec:C-in-QM-B} we show that, for a weak-measurement protocol, the correlator of the pointer readouts tends in the weak limit to the same symmetrized correlator when $Q$ is not necessarily dichotomic \cite{Wang2002}.

\section{Leggett-Garg inequality \label{ssec:LGI}}
The LGI in Eq.\ (2) of the main text concerns the two-time correlation functions
\begin{equation} \label{seq:Cij-def}
C_{ij} = \braket{q(t_i)q(t_j)} = \sum_{q_i, q_j} q_i q_j \,p_{ij}(q_i, q_j)
\end{equation}
of an observable $q$. Here, $p_{ij}(q_i, q_j)$ denotes the joint probability of obtaining outcomes $q_i$ and $q_j$ from two successive measurements of $q$ performed at times $t_i$ and $t_j$, respectively, when only these two measurements are performed. Typically, $q$ is taken to be dichotomic, having only two possible measurement results $\pm 1$. However, the LGI applies also to observables with more than two (possibly infinitely many) possible outcomes, provided $|q| \le 1$. The LGI states that \cite{Leggett:1985zz, Emary:2013wfl}, for any initial state of the system,
\begin{equation} \label{seq:LGI}
    K \le 1, \qquad \qquad K \equiv C_{12} + C_{23} - C_{13},
\end{equation}
for arbitrary times $t_1$, $t_2$, and $t_3$.

As we now show, Eq.~\eqref{seq:LGI} follows directly from two assumptions: (i) $q$ has a definite value at all times, regardless of whether it is measured, and (ii) measurements of $q$ can be performed with negligible effect on subsequent dynamics. Assumption (i) is known as \emph{realism per se}, and assumption (ii) is called \emph{noninvasive measurability}. By assumption (i), when measurements are made at $t_i$ and $t_j$, $q$ nevertheless has a definite value at time $t_k \neq t_i, t_j$. Therefore, there exists an underlying distribution $p_{ij}(q_1, q_2, q_3)$ giving the probability that the values of $q$ are $q_1$, $q_2$, and $q_3$ at times $t_1$, $t_2$, $t_3$, despite measurements occurring only at $t_i$ and $t_j$. The pairwise distributions $p_{ij}(q_i, q_j)$ are then obtained by marginalizing:
\begin{equation} \label{seq:marginalization}
\begin{aligned}
    p_{12}(q_1, q_2) = \sum_{q_3}  p_{12}(q_1, q_2, q_3),
\end{aligned}
\end{equation}
etc., where $\sum_{q_3}$ denotes a sum and/or integral over possible values of $q$. Additionally, by assumption (ii), $p_{ij}(q_1, q_2, q_3)$ does not depend on $t_i$ and $t_j$, so we write $p_{ij}(q_1, q_2, q_3) = p(q_1, q_2, q_3)$. Then, combining Eqs.~\eqref{seq:Cij-def} and \eqref{seq:marginalization}, we have
\begin{align} \label{seq:K-as-sum}
    K = \sum_{q_1,q_2,q_3} \left[ q_1q_2 + q_2q_3 - q_1q_3 \right]p(q_1,q_2,q_3).
\end{align}
Finally, we insert the termwise upper bound
\begin{equation}\label{seq:LGI-termwise-bound}
    \begin{aligned}
        q_1q_2 + q_2q_3 - q_1q_3 &= q_2(q_1+q_3)-q_1q_3 \\
        &\le |q_1+q_3| - q_1 q_3 \\
        &\le 1,
    \end{aligned}
\end{equation}
which follows from the assumption $|q_i|\le 1$, into Eq.~\eqref{seq:K-as-sum}, establishing the LGI of Eq.~\eqref{seq:LGI}.

\section{Quantum-mechanical calculation of $C_{ij}$ \label{ssec:C-in-QM}}

In a quantum system, the observable $q$ corresponds to a Hermitian operator $Q$. In general, the operator expression for $C_{ij}$, defined in Eq.~\eqref{seq:Cij-def}, depends on the measurement protocol. As we show below, for dichotomic observables measured projectively, $C_{ij}$ is given by the symmetrized correlation function \cite{Fritz:2010qzm}
\begin{equation} \label{seq:Cij-symmetrized}
    \frac{1}{2} \Tr [ \rho_0 \{ Q(t_i), Q(t_j)\}],
\end{equation}
where $\{\cdot,\cdot\}$ denotes the anticommutator, $ \rho_0$ is the initial state at time $t=0$, and $ Q(t) = e^{i  H t} Q e^{-i  H t}$. For non-dichotomic but bounded observables, we will instead adopt a weak-measurement protocol. In an ideal macrorealist, noninvasive detector model, discussed explicitly below, the correlator of the detector outputs reproduces the classical correlator of Eq.~\eqref{seq:Cij-def}, while in quantum mechanics the same protocol yields Eq.~\eqref{seq:Cij-symmetrized} in the weak limit \cite{Wang2002}.

\subsection{Dichotomic case \label{ssec:C-in-QM-A}}

In the case of a dichotomic observable $Q$ such that $Q^2 = 1$, we adopt a projective measurement protocol \cite{Fritz:2010qzm}, which proceeds as follows in the Schr\"odinger picture. Let us take $i=1$ and $j=2$ for concreteness. First, $ \rho_0$ evolves unitarily into $ \rho_1 = e^{-i  H t_1} \rho_0e^{i  H t_1}$ under the Hamiltonian $H$. Then, a projective measurement of $ Q$ is carried out at time $t_1$. By the Born rule, the probability of obtaining outcome $q_1 = \pm 1$ is $p_{1}(q_1) = \Tr[ \rho_1 \Pi(q_1)]$, where $ \Pi(q_1) = \frac{1}{2}\left(1 + q_1  Q\right)$ projects onto the $q_1=\pm 1$ eigenspace of $Q$. After obtaining the outcome $q_1$, the state collapses to $ \rho_1' =  \Pi(q_1)  \rho_1  \Pi(q_1)/p_1(q_1)$. This state evolves unitarily to $ \rho_2 = e^{-i  H (t_2-t_1)}  \rho_1' e^{i  H (t_2-t_1)}$, whereupon $ Q$ is measured again, giving outcome $q_2$ with probability conditioned on the outcome $q_1$
\begin{equation}
    p_{12}(q_2 | q_1) = \Tr\left[  \rho_2 \Pi(q_2) \right] = \frac{1}{p_1(q_1)}\Tr\left[ e^{-i  H (t_2-t_1)}  \Pi(q_1) e^{-i  H t_1} \rho_0 e^{i  H t_1}  \Pi(q_1) e^{i  H (t_2-t_1)}  \Pi(q_2) \right].
\end{equation}
The joint probability of obtaining $q_1$ followed by $q_2$ is therefore $p_{12}(q_1,q_2) = p_1(q_1) p_{12}(q_2|q_1)$, and the correlation function \eqref{seq:Cij-def} becomes
\begin{equation}
    C_{12} = \sum_{q_1,q_2} q_1 q_2 \Tr\left[ e^{-i  H (t_2-t_1)}  \Pi(q_1) e^{-i  H t_1}  \rho_0 e^{i  H t_1}  \Pi(q_1) e^{i  H (t_2-t_1)}  \Pi(q_2) \right].
\end{equation}
Using cyclicity of the trace and collecting $e^{i  H t_2}\left( \sum_{q_2} q_2  \Pi(q_2) \right)e^{-i  H t_2} =  Q(t_2)$, we have
\begin{equation}
\begin{aligned}
    C_{12} &= \sum_{q_1 = \pm 1} q_1\Tr\left[  Q(t_2) e^{i  H t_1}  \Pi(q_1) e^{- i  H t_1} \rho_0 e^{i  H t_1}  \Pi(q_1)e^{-i  H t_1} \right] \\
    &= \frac{1}{4}\sum_{q_1 = \pm 1} q_1\Tr\left[  Q(t_2) e^{i  H t_1}\left( 1 + q_1  Q \right) e^{- i  H t_1} \rho_0 e^{i  H t_1} \left( 1 + q_1  Q \right)e^{- i  H t_1} \right].
\end{aligned}
\end{equation}
Recognizing that $e^{i  H t_1}\left( 1 + q_1  Q\right) e^{-i  H t_1} = 1 + q_1 Q(t_1)$ and using the dichotomic property $[ Q(t_1)]^2 = 1$ and $\sum_{q_1} q_1 = 0$, this simplifies to the symmetrized correlator
\begin{align}
    C_{12} &= \frac{1}{2} \Tr[\rho_0\{Q(t_1),Q(t_2)\}],
\end{align}
as in Ref.~\cite{Fritz:2010qzm}.

\subsection{Non-dichotomic case \label{ssec:C-in-QM-B}}

The argument just given for the dichotomic case fails for non-dichotomic $Q$, for which $Q^2 \neq 1$, because the projector $\Pi(q)$ no longer has a simple expression in terms of $Q$. However, for the weak-measurement protocol described below, the corresponding meter-output correlator tends in the weak limit to Eq.~\eqref{seq:Cij-symmetrized} \cite{Wang2002}.

First, let us review the standard von Neumann measurement procedure \cite{Von_Neumann2018-ez, Preskill:QInotes}. Suppose we want to measure the observable $Q$ in a system of interest. The measurement is carried out by coupling the system to a measurement device (the ``pointer''), described by an interaction Hamiltonian
\begin{align}
     H_\text{int}(t) = - g(t) \lambda\, Q \otimes  X,
\end{align}
where $X$ is the canonical variable of the pointer with conjugate momentum $P$, i.e., $[X, P]=i$. We assume the function $g(t)$ satisfies $\int_{-\infty}^\infty dt\ g(t) = \int_{0}^{\Delta t}dt\ g(t) = 1$ and consider the impulsive limit in which the interaction duration $\Delta t$ is short enough that the intrinsic dynamics of the system and pointer can be neglected during the interaction. Unitary evolution to time $\Delta t$ is thus provided by $U = e^{i \lambda Q \otimes X}$, and we have
\begin{align} \label{seq:PDeltat}
    P(\Delta t) = U^\dag P U = P + \lambda Q,
\end{align}
or $Q = \delta P/\lambda$, where $\delta P = P(\Delta t) - P(0)$. Hence, if the pointer is initially prepared in a momentum eigenstate, then projectively measuring its momentum after the interaction effects a measurement of $Q$.

Suppose that instead of a momentum eigenstate, the pre-interaction state $\ket{\psi_0}$ of the pointer is a minimum-uncertainty state
\begin{equation} \label{seq:phi-tminus}
\begin{aligned}
    \ket{\psi_{0}} &= \frac{1}{\sqrt{\Delta X \sqrt{2\pi}}}\int dX\ \exp\left( - \frac{X^2}{4(\Delta X)^2} \right) \ket{X} \\
    &= \sqrt{\sqrt{\frac{2}{\pi}}\Delta X}\int dP\  e^{-(\Delta X)^2P^2}\ket{P},
\end{aligned}
\end{equation}
where the normalizations $\braket{X|X'} = \delta(X-X')$ and $\braket{P|P'} = \delta(P-P')$ are chosen, so that $\braket{\psi_0|\psi_0} = 1$. We write the state of the combined system just prior to the measurement procedure as $ \rho_{0} =  \sigma_{0}\otimes \ketbra{\psi_{0}}{\psi_{0}}$, where $ \sigma_{0} = \sum_{nm} \bra{q_n} \sigma_{0}\ket{q_m} \ketbra{q_n}{q_m}$ in the eigenbasis $\{\ket{q_n}\}$ of $Q$. Then, in the Schr\"odinger picture, the post-interaction state is
\begin{align} \label{seq:rho-t-plus}
     \rho(\Delta t) = \sum_{nm} \bra{q_n} \sigma_{0}\ket{q_m}\ketbra{q_n}{q_m} \otimes \ketbra{\psi_{q_n}}{\psi_{q_m}},
\end{align}
where
\begin{equation}
\begin{aligned} \label{seq:phi-tplus}
    \ket{\psi_{q_n}} &= e^{i\lambda q_n  X}\ket{\psi_{0}} \\
    &= \sqrt{\sqrt{\frac{2}{\pi}}\Delta X} \int dP\ e^{-(\Delta X)^2 P^2}\ket{P+\lambda q_n}.
\end{aligned}
\end{equation}
The probability density $p(P)$ that a measurement of the pointer's momentum yields value $P$ is
\begin{align} \label{seq:pointer-distribution}
    p(P) = \Tr[\rho(\Delta t) \ketbra{P}{P}] = \sqrt{\frac{2}{\pi}}\Delta X \sum_n \braket{q_n|\sigma_0|q_n} e^{-2(\Delta X)^2(P-\lambda q_n)^2},
\end{align}
and the (unnormalized) post-measurement state of the system is
\begin{align}
    \sigma_\text{post} = \Tr_\text{pointer}\left[ \ketbra{P}{P} \rho(\Delta t) \ketbra{P}{P} \right] \propto \sum_{n,m} \braket{q_n|\sigma_0|q_m} e^{-(\Delta X)^2 \left[ (P-\lambda q_n)^2 + (P- \lambda q_m)^2 \right]}\ketbra{q_n}{q_m}.
\end{align}
In the limit $\Delta X \rightarrow \infty$, we have $p(P) \rightarrow \sum_{n}\braket{q_n|\sigma_0|q_n} \delta(P - \lambda q_n)$, i.e., the possible outcomes of measuring the pointer's momentum are $P = \lambda q_n$, occurring with probability $\braket{q_n|\sigma_0|q_n}$ (assuming $q_n$ is non-degenerate). The corresponding post-measurement state is then
\begin{align}
    \sigma_\text{post} \rightarrow \ketbra{q_n}{q_n},
\end{align}
recovering the projective measurement of $Q$ in state $\sigma_0$.

In the opposite limit $\Delta X \rightarrow 0$, the distribution $p(P)$ becomes uniform, and $\sigma_\text{post} \rightarrow \sigma_0$. Thus, one learns vanishingly little about the state of the system while also perturbing it negligibly. This is the weak measurement regime \cite{Aharonov:1988xu, Aharonov:1990zza}. Nevertheless, even for small, finite $\Delta X$, measurements of $P$ on many identical preparations of the system can be used to reconstruct the expectation value $\braket{Q}_0 = \Tr\left[ \sigma_0 Q\right]$. Indeed, using $\rho(\Delta t) = U \rho_0 U^\dag$ and Eq.~\eqref{seq:PDeltat}, we have
\begin{equation}
\begin{aligned}
    \braket{P}_{\Delta t} &= \Tr\left[ \rho(\Delta t) P \right] \\
    &= \Tr[\rho_0 U^\dag P U] \\
    &= \braket{P}_0 + \lambda \braket{Q}_0,
\end{aligned}
\end{equation}
or
\begin{align} \label{seq:one-time}
    \braket{Q}_0 = \frac{\braket{P}_{\Delta t} - \braket{P}_0}{\lambda} = \frac{\braket{P}_{\Delta t}}{\lambda},
\end{align}
since $\braket{P}_0 = 0$.

\subsubsection{The two-time correlation function}

Now consider measuring $Q$ twice on the same system. To do so, we couple the system sequentially to two independent pointers $(X_1, P_1)$ and $(X_2, P_2)$, both prepared in identical minimum-uncertainty states $\ket{\psi_0}$ as in \eqref{seq:phi-tminus}. The state of the full system is $\rho_0 = \sigma_0 \otimes \ketbra{\psi_0}{\psi_0}_1 \otimes \ketbra{\psi_0}{\psi_0}_2$. The first interaction, performed at time $t_1=0$ for simplicity, couples the system to pointer 1 via $U_1 = e^{i\lambda Q \otimes X_1}$. Then, the combined system evolves freely under its full Hamiltonian $H = H_\text{sys} + H_1(P_1) + H_2(P_2)$ for a time $\tau$, where the three terms act on different Hilbert spaces. We assume that $H_1$ and $H_2$, the Hamiltonians for the pointers, have the same form and are independent of $X_1$ and $X_2$, respectively, e.g.~$H_i = P_i^2/2m$. At time $t_2=\tau$, the second interaction couples the system to pointer 2 via $U_2 = e^{i\lambda Q \otimes X_2}$. The full time-evolution operator is therefore $U(\tau) = U_2\, e^{-iH\tau}\, U_1$, and the post-interaction state is $\rho(\tau) = U(\tau) \rho_0 U^\dagger(\tau)$.

The measurement of pointer 1 can be made immediately after the interaction between the system and pointer 1, or it can be made at any later time, since pointer 1 decouples from the system after the interaction and $P_1$ commutes with $H_1(P_1)$. For simplicity, we choose to measure both pointers simultaneously after the final interaction at time $\tau$.

It is important to distinguish the bounded variable $q$ appearing in the LGI derivation of Sec.~\ref{ssec:LGI} from the raw pointer readouts. In weak-measurement LGI implementations, one works operationally with noisy detector outputs rather than ideal projective values of the system variable \cite{Jordan:2006lrf,Emary:2013wfl}. In the present weak-measurement protocol, the rescaled pointer outcomes
\begin{equation}
    m_1 \equiv \frac{P_1}{\lambda}, \qquad m_2 \equiv \frac{P_2}{\lambda}
\end{equation}
serve as noisy estimators of the underlying bounded observable $q$; they should not be identified literally with the values $q_1$ and $q_2$ of Sec.~\ref{ssec:LGI}, since the pointer outcomes are in general unbounded, as seen from Eq.~\eqref{seq:pointer-distribution}. The weak protocol therefore defines the experimentally accessible pointer-output correlator
\begin{align} \label{seq:C12-two-time}
    \widetilde C_{12}
    \equiv
    \langle m_1 m_2 \rangle_\tau
    =
    \left\langle\frac{P_1}{\lambda} \otimes \frac{P_2}{\lambda}\right\rangle_\tau
    =
    \frac{1}{\lambda^2}\Tr[\rho(\tau) P_1 \otimes P_2].
\end{align}

Let us now compute $\braket{P_1 \otimes P_2}_\tau \equiv \Tr[\rho(\tau)\, P_1 \otimes P_2]$ using cyclicity of the trace:
\begin{equation}
\begin{aligned}
    \braket{P_1 \otimes P_2}_\tau &= \Tr[\rho_0\, U(\tau)^\dagger (P_1 \otimes P_2)\, U(\tau)] \\
    &= \Tr[\rho_0\, U_1^\dagger\, e^{iH\tau}\, U_2^\dagger\, (P_1 \otimes P_2)\, U_2\, e^{-iH\tau}\, U_1].
\end{aligned}
\end{equation}
Since  $U_2$ acts trivially on pointer 1, and both $P_1$ and $P_2$ are constants of motion for $H$, we have
\begin{align}
    U_1^\dagger\, e^{iH\tau}\, U_2^\dagger\, (P_1 \otimes P_2)\, U_2\, e^{-iH\tau}\, U_1 = U_1^\dag \left[ P_1 \otimes (P_2 + \lambda Q(\tau))\right] U_1,
\end{align}
where $Q(\tau) = e^{iH\tau} Q\, e^{-iH\tau}$. Performing the trace over pointer 2 yields
\begin{align}
    \braket{P_1 \otimes P_2}_\tau = \lambda \Tr\left[ \sigma_0 \otimes \ketbra{\psi_0}{\psi_0}_1 U_1^\dag \left[ Q(\tau) \otimes P_1\right]U_1\right].
\end{align}
Inserting $U_1 = \sum_n e^{i \lambda q_n X_1} \ketbra{q_n}{q_n}$, performing the trace, and dropping the subscript $1$, we find
\begin{align}
    \braket{P_1 \otimes P_2}_\tau = \lambda \sum_{n, m} \braket{q_n|Q(\tau)|q_m} \braket{q_m|\sigma_0|q_n} \braket{\psi_0|e^{-i \lambda(q_n-q_m)X}(P + \lambda q_m)|\psi_0}.
\end{align}
From Eq.~\eqref{seq:phi-tminus}, we compute
\begin{align}
    \braket{\psi_0|e^{-i a X}|\psi_0} = e^{- a^2 (\Delta X)^2/2}, \qquad \qquad \braket{\psi_0|e^{-i a X}P|\psi_0} = \frac{a}{2} e^{- a^2 (\Delta X)^2/2}.
\end{align}
Therefore, in the weak measurement limit $\Delta X \rightarrow 0$,
\begin{equation}
\begin{aligned}
    \braket{P_1 \otimes P_2}_\tau &\rightarrow \frac{\lambda^2}{2} \sum_{n, m} \braket{q_n|Q(\tau)|q_m} \braket{q_m|\sigma_0|q_n} \left(q_n + q_m \right) \\
    &= \frac{\lambda^2}{2} \braket{\{Q(\tau),Q\}}_0,
\end{aligned}
\end{equation}
or
\begin{align}
    \left\langle\frac{P_1}{\lambda} \otimes \frac{P_2}{\lambda}\right\rangle_\tau \rightarrow \frac{1}{2}\braket{\{Q(\tau),Q\}}_0.
\end{align}
This was first shown in Ref.~\cite{Wang2002}. By comparison with Eq.~\eqref{seq:C12-two-time}, we conclude that in the weak limit
\begin{equation}\label{seq:quantum-two-time}
\begin{aligned}
    \widetilde C_{12}
    \rightarrow
    \frac{1}{2}\braket{\{Q(\tau),Q\}}_0.
\end{aligned}
\end{equation}

To connect $\widetilde C_{12}$ with the classical correlator $C_{12}$ entering the LGI of Sec.~\ref{ssec:LGI}, consider the corresponding macrorealist, noninvasive description of the same weak-measurement protocol. Inspired by the ambiguous-detector viewpoint reviewed in Ref.~\cite{Emary:2013wfl} and by the signal-plus-noise weak-detector model used in Ref.~\cite{Williams:2007cnr}, we model the pointer outputs as idealized unbiased noisy readouts of the hypothetical definite (in the sense of \emph{realism per se}, as defined in Sec.~\ref{ssec:LGI}) bounded values $q(0)$ and $q(\tau)$, namely
\begin{equation}
    m_1 = q(0) + \xi_1, \qquad m_2 = q(\tau) + \xi_2,
\end{equation}
where $\xi_1$ and $\xi_2$ are detector noises with zero mean, uncorrelated with the system and with each other at distinct times. In this idealized detector model, the weak and noninvasive regime is represented by taking the variances of $\xi_1$ and $\xi_2$ to be large. Under this detector model, the correlator of pointer outputs is
\begin{equation} \label{seq:classical-two-time}
\begin{aligned}
    \widetilde C_{12}
    = \langle m_1 m_2 \rangle
    &= \langle q(0)q(\tau)\rangle
    + \langle q(0)\xi_2\rangle
    + \langle \xi_1 q(\tau)\rangle
    + \langle \xi_1\xi_2\rangle \\
    &= \langle q(0)q(\tau)\rangle
    = C_{12}.
\end{aligned}
\end{equation}
Thus, in this macrorealist, noninvasive detector model, the pointer-output correlator $\widetilde C_{12}$ reproduces the very same classical correlator of $q(0)$ and $q(\tau)$ that enters the proof of the LGI. From a macrorealist perspective, one therefore expects the experimentally determined Leggett-Garg combination
\begin{equation}
\widetilde K \equiv \widetilde C_{12} + \widetilde C_{23} - \widetilde C_{13}
\end{equation}
to satisfy $\widetilde K \le 1$. On the other hand, from the quantum-mechanical perspective, the experimentally determined $\widetilde C_{ij}$ tends to the corresponding symmetrized quantum correlators of $Q$ in the weak-measurement limit, and the latter can produce a $\widetilde K$ exceeding unity. In this operational sense, the weak-measurement protocol provides the appropriate implementation of the LGI test for non-dichotomic variables \cite{Goggin:2009rae,Emary:2017jtt}.

\bibliography{refs}